\RequirePackage{fix-cm}
\documentclass[smallextended,twocolumn]{svjour3}       
\smartqed  
\usepackage{graphicx}
\usepackage{color}

\journalname{Journal of Superconductivity and Novel Magnetism}
\begin{document}
\onecolumn
\title{Mass enhancements and band shifts in strongly hole overdoped 
Fe-based pnictide superconductors: KFe$_2$As$_2$ and CsFe$_2$As$_2$
}


\onecolumn
\author{S.-L.\ Drechsler$^1$, H.~Rosner$^2$, V.\ Grinenko$^{1,3}$, S.~Aswartham$^1$,
I.~Morozov$^{1,4,10}$, M.\ Liu$^{1,4}$, A.\ Boltalin$^4$, 
K.~Kihou$^5$,\\
 C.H.~Lee$^5$, T.K.\ Kim$^6$, D.~Evtushinsky$^7$, 
J.M.\ Tomczak$^8$, S.\ Johnston$^9$, S.~Borisenko$^1$}

\authorrunning{S.-L.\ Drechsler, S.\ Rosner, V.\ Grinenko {\it et al.} } 

\institute{$^1$ Leibniz Inst.\ of Solid State Research  \& Material Science \at IFW-Dresden
              D-01168 Dresden, 
              Helmholtzstr.\ 20
              Germany\\ 
              Tel.: +49-351-4659384, Fax: +49-351-4659400\\
              \email{s.l.drechsler@ifw-dresden.de} \\
$^2$Max-Planck-Institute for  Chemical Physics of Solids \at 
D-01187 Dresden, Germany\\
$^3$ Institute of Solid State and Materials Physics, \at TU Dresden,  D-01169 Dresden, Germany\\
$^4$  Lomonosov Moscow State University, \at Department of Chemistry, 119991 Moscow, Russia\\
$^5$ Nat.\ Inst.\ of Advanced Industrial Science \& Technology, \at (AIST)
Tsukuba, Ibaraki 305-8568, Japan\\
$^6$Diamond Light Source, Harvell Science \& Innovation Camp. \at 
Didcot OX110DE, United Kingdom\\
$^7$Inst. of Physics, Ecole Polytechnique Federale Lausanne  \at  CH-1015 Lausanne, Switzerland\\
$^8$ Vienna-University of Technology, \at Institute of Solid State Physics, A-1040  Wien, Austria\\
$^9$Department of Physics and Astronomy, Tennesee University, \at Tennesee 37996  Knoxville, USA\\
$^{10}$Lebedev Physical Institute, Russian Academy of Science, 11991 Moscow, Russia
}

          
\vspace{-3cm}
\date{\small Received: 14 November 2017 / Accepted: 15 November 2017}

\maketitle

\begin{abstract}
The interplay of high and low-energy mass renormalizations with 
band-shifts reflected by the positions of van Hove singularities (VHS)
in the normal state spectra
of the highest
hole-overdoped and
strongly  
correlated 
AFe$_2$As$_2$ (A122) 
with A=K, Cs is discussed phenomenologically based
on ARPES data and GGA band-structure calculations with full
spin-orbit coupling. 
The big increase
of the Sommerfeld coefficient $\gamma$ from K122 to Cs122 is ascribed
to an enhanced coupling to low-energy bosons
in the vicinity of a quantum critical point to
an  unknown, yet incommensurate phase  
different from the commensurate Mott one.
We find no sizeable increase in correlations for Cs122
in contrast to F.\ Eilers {\it et al.}, PRL {\bf 116}, 237003 (2016) [3].
The empirical (ARPES) VHS positions as
compared with GGA-predictions 
point even to slightly weaker correlations in Cs122 in accord with 
low-$T$ magnetic susceptibility $\chi(T)$ data and a decreasing Wilson ratio
$\propto \chi(0)/\gamma$.
\end{abstract}
\begin{figure}[t]
  \includegraphics[width= 
  1.04\textwidth]{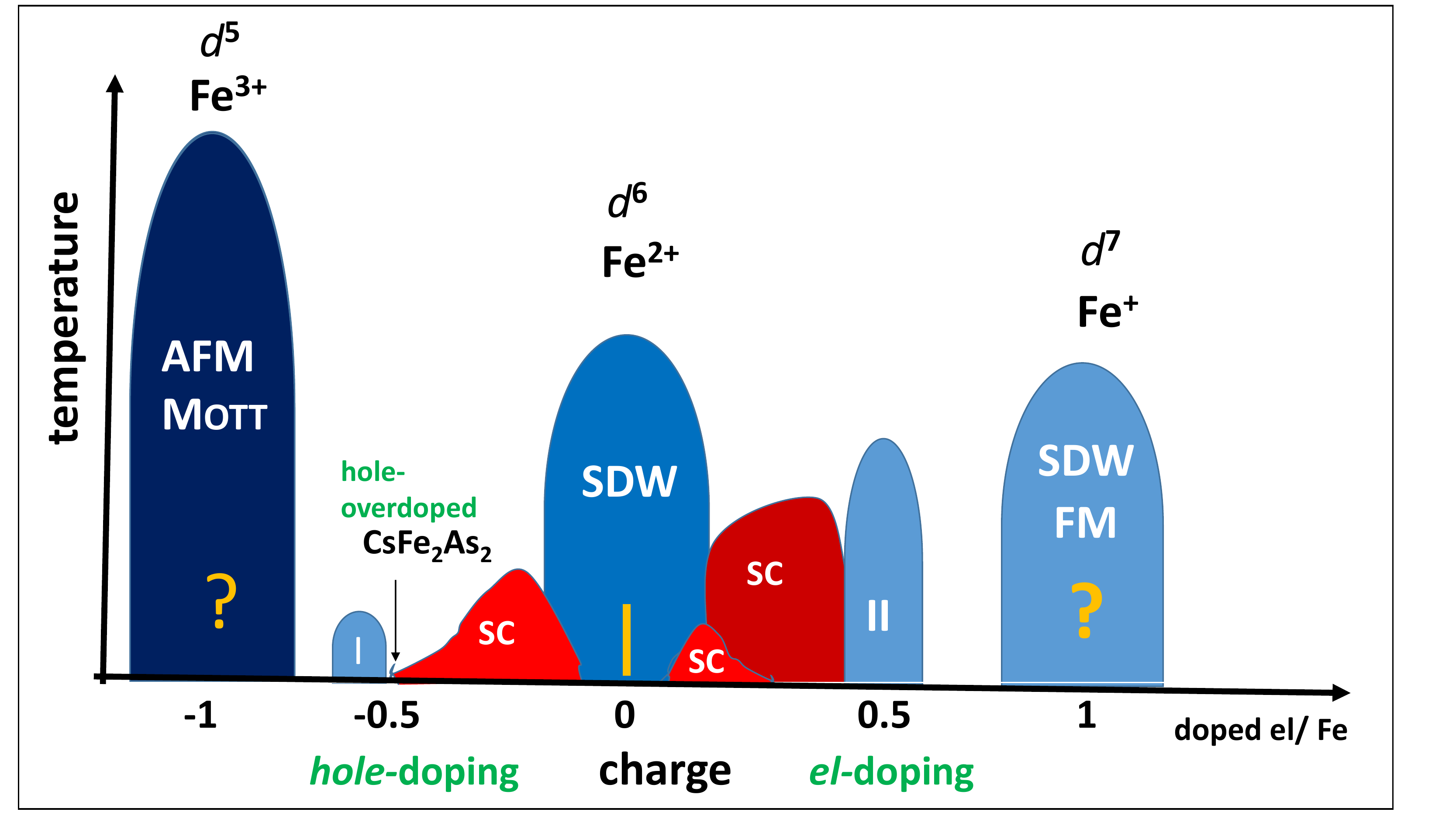}
\caption{(Color) Schematic general doping phase diagram of 
Fe pnictides ignoring nematicity.
Blue (red) magnetic (superconducting) regions, respectively. Phase I - a combined
charge, orbital, and spin ordered phase responsible for the vicinity of the critical
point as discussed in the text. The yellow line at isovalent or no doping
stands for such systems as Li(Na)FeAs and P-doped Ba(Sr)-122 where
the competing magnetic SDW magnetic stripe-phase is absent or strongly suppressed.
Phase II has been observed but not been yet characterized experimentally.
The outermost hypothetical SDW or ferromagnetic (FM) phase around Fe$^+$ 
is our suggestion.
The bright (dark) red regions stand for 122 and H doped La-1111
(under pressure) \cite{Kawaguchi2016}
FeSC  compounds, respectively. 
}
\label{fig:1}       
\end{figure}
\begin{figure}[b]
  \includegraphics[width=1.02\textwidth]{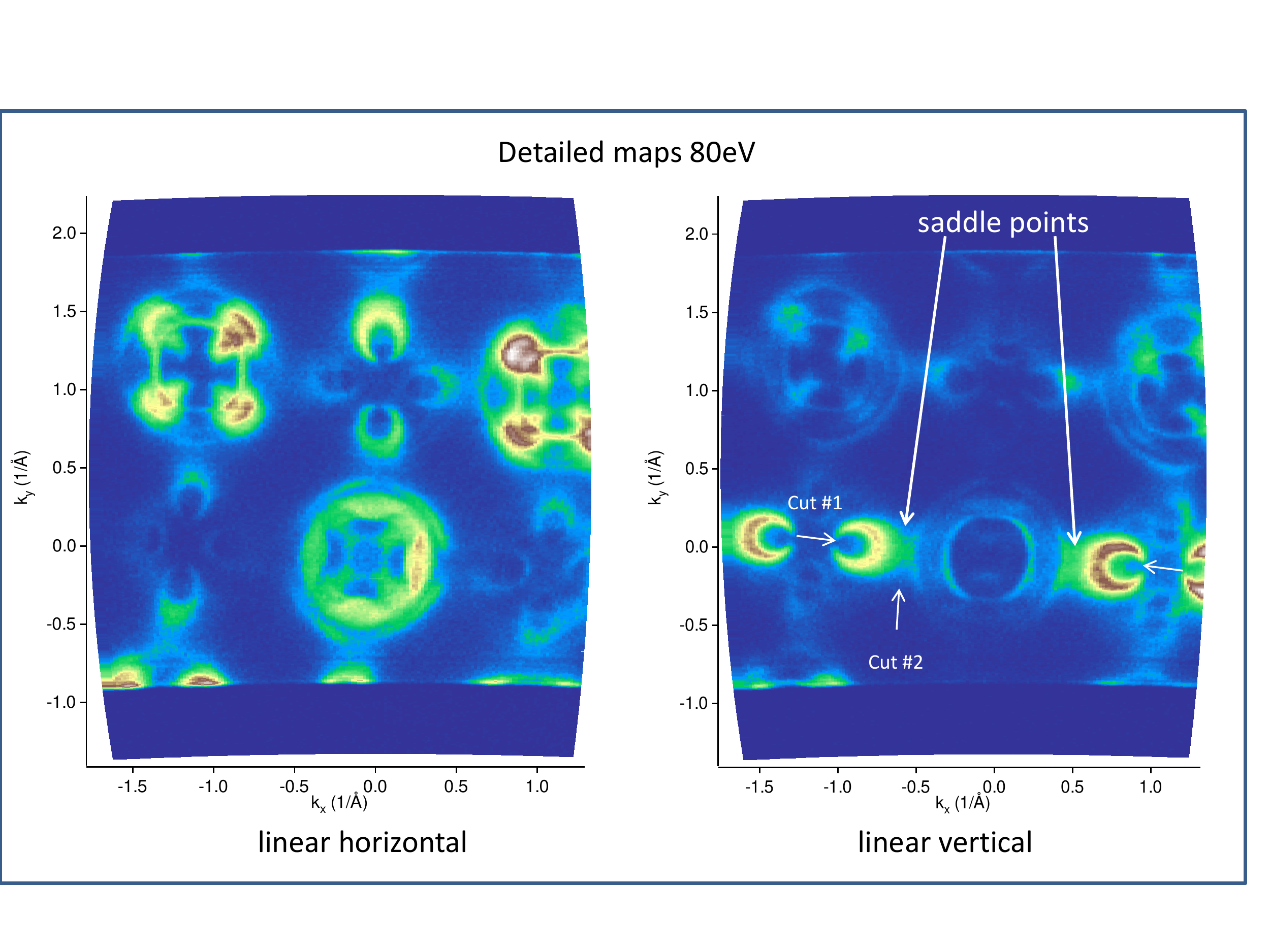}
  \includegraphics[width=1.02\textwidth]{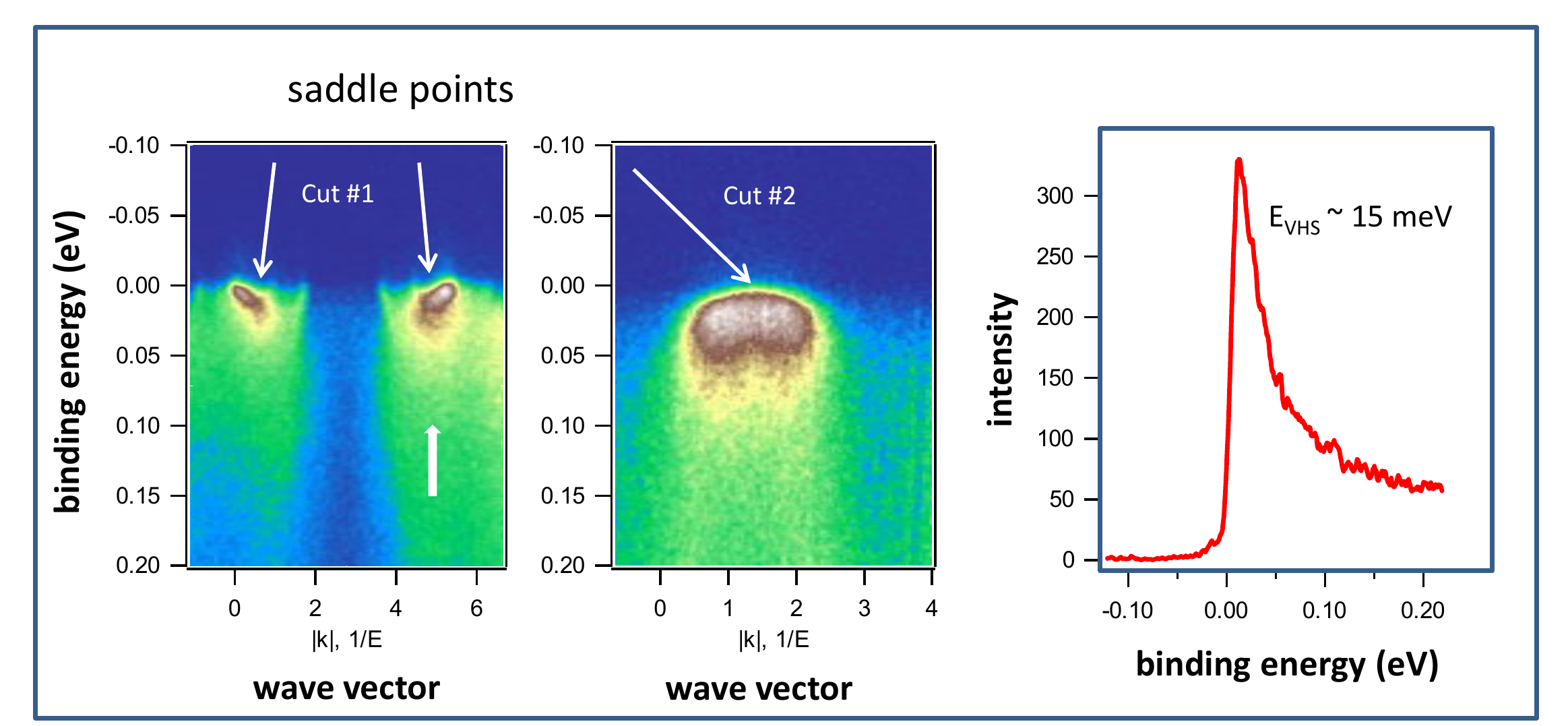}
\caption{(Color) ARPES data for K$_{0.96}$Cs$_{0.04}$Fe$_2$As$_2$.
Upper panel:general map with visible $\varepsilon$-FSS,
Lower panel: Cuts (white arrows) for the linearly polarized
spectra (left and middle); 3$d_{xz}/3d_{yz}$ derived saddle-point  
VHS (right). }
\label{fig:7}       
\end{figure} 
\twocolumn
\section{Introduction}
\label{intro}
To achieve a better
qualitative and quantitative understanding
of the rich doping phase diagram and
the strength of correlation effects
in the electronic and magnetic properties
of Fe-based superconductors and related compounds
is one of the most challenging and central issues
(see e.g.\  Fig.\ 1 and \cite{Drechsler2017}).
The two isomorphic compounds $A$Fe$_2$As$_2$ with $A=$~K, Cs
with the same strongly hole-overdoped state 
but with surprisingly different thermodynamic properties  
in the vicinity of a recently proposed quantum critical point (QCP)
\cite{Eilers2016}
provide a good opportunity 
to get deeper insight into the interplay
of high- and low-energy physics by a comparative
analysis of the underlying electronic structure:
In Sects.\ 3-5 we present densities of states (DOS)
around the Fermi level and discuss band shifts as marked by the positions of 
van Hove singularities (VHS). The very presence  of VHS for one of them was first qualitatively 
detected in STM%
\footnote{ (Scanning tunnelling measurements). It includes in our opinion, however, an artificial
upshift of about 10~meV of the entire ARPES spectrum, see also Section 5.}
on K122 surfaces \cite{Fang2015} (its shape is similar to that shown Figs.\ 2-4). The same group
did, however, not succeed in observing an expected analogous feature in Cs122\cite{Yang2016}. 
From the theoretical DFT side a clear identification of the VHS
was previously lost due to a too scarce k-mesh\cite{Drechsler2017}.
In Sect.\ 5 we report the observation of VHS for both
compounds by means of ARPES. 
Controversial aspects of the magnetic susceptibilities,  $\chi(T)$,
and the Wilson ratios of the title compounds are discussed
in Sect.\ 4.
The present paper builds on the recent
multiband Eliashberg theory based analysis
of Fe-based superconductors
of
Refs.\ \cite{Drechsler2017,Efremov2017} supplemented, here, with
new ARPES and thermodynamic
experimental findings
for $A$122 ($A=$~K, Cs) including also the less strongly K-overdoped 
Ba$_{1-x}$K$_x$Fe$_2$As$_2$
system with $0.7 \leq x < 1$. 
\section{Sample preparation and characterization}
\label{sec:2}
For the ARPES (angular resolved photoemission spectroscopy) measurements 
reported in Sect.\ 5 (see Fig.\ 2 )
 K$_{0.96}$Cs$_{0.04}$ and 
 Cs$_{0.94}$K$_{0.06}$Fe$_2$As$_2$
 single crystals 
have been grown from corresponding $A$As ($A=$K, Cs) mixed fluxes
\cite{Hafiez2012,Hafiez2013,Grinenko2014,Kihou2016}.
The electronic structure and the normal state properties of both samples, 
 we focus here, 
 are considered to be very close to those of the 
 corresponding stoichiometric cases.
Both as grown crystals exhibit a plate-like morphology, these crystals were characterized 
by SEM (scanning electron microscopy) and
EDX (energy-dispersive X-ray analysis) diffraction
measurements. 
The weak disorder produced by the slight deviations from stoichiometry
causes however an expected and observed reduction of $T_c$ generic for
unconventional nodal superconductivity, 
similarly as in the case of Na-doped K122 \cite{Efremov2017}
generic for an $d$-wave superconducting order parameter.
Also in the cases of less strongly hole-overdoped samples
Ba$_{1-x}$K$_x$Fe$_2$As$_2$ shown in Figs.\ 6 and 7 the 
self-flux
method for the crystal growth \cite{Kihou2016}
have been used resulting in high-quality single crystals with 
low values for the residual resistivity 
$\rho_0$. The doping level was determined by x-ray diffraction using the known 
dependence of the $c$-axis lattice
constant versus K doping (see Fig.\ 3 in Ref.\ 9 and Fig.\
S1 in the Suppl.\ of Ref.\ 8). 
The specific heat and electric transport measurements of single crystals 
were done in a Quantum Design physical property measurement system (PPMS) 
and the magnetization measurements in a commercial superconducting 
quantum interference device magnetometer (SQUID) from Quantum Design. 
Part of the magnetization, transport and specific heat data shown here
have been presented already in Ref. \cite{Grinenko2017}.

\section{Mass enhancement from the density of states}
\begin{figure}[t]
  \includegraphics[width=0.42\textwidth]{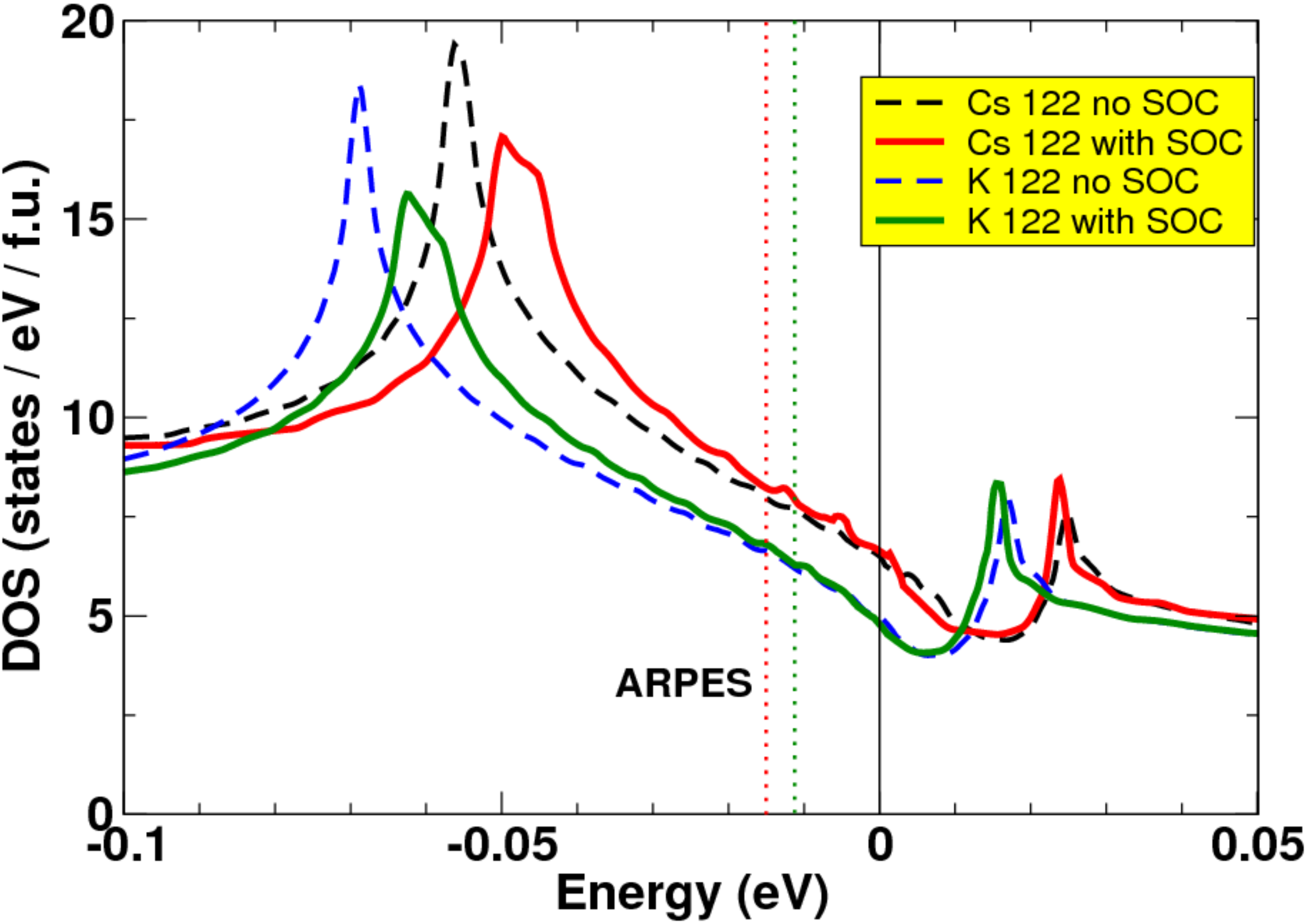}
\caption{(Color online) Bare densities of states (DOS) for K122 (green) and Cs122 (red) from
GGA calculations with (full lines) and without (dashed lines)  spin orbit coupling (SOC) included. 
Thin dotted lines - experimental (ARPES) positions of the 3$d_{xz}/3d_{yz}$ van Hove
singularities (see also text.)}
\label{fig:2}       
\end{figure}
Here we will consider  the electronic DOS, $N(0)$,
 at the Fermi energy $\varepsilon_{\rm \tiny F}=0$
and  several related thermodynamic
properties such as the Sommerfeld coefficient $\gamma$ of the linear
in temperature $T$ electronic specific heat at low $T$ (see Fig.\ 4),
the magnetic susceptibilities $\chi(T)$ (see Figs.\ 5 and 6),
 as well as the positions of
saddle-point VHS generic
for quasi-2D electronic systems.
In the present case, these derive from Fe-3$d_{xz}/d_{yz}$
electronic states and, owing to the strong hole-doping  level
occurs only slightly below $\varepsilon_{\rm \tiny F}$  (see Figs.\ 2, 3, and 4),
as compared to the parent compounds $A$Fe$_2$As$_2$  with
$A=$~Ba, Sr, Ca or isovalent P-doped Ba122 or Sr122 (see Fig.\ 1)
with Fe$^{2+}(d^6)$ .
However, all these quantities are strongly
affected by the presence of non-negligible many-body effects
which become evident when
compared with traditional density functional theory 
(DFT) based results, within the framework of the local density
approximation (LDA) and the general gradient approximation (GGA)
(for details see Ref.\ \cite{Drechsler2017}) 
employed here as an effective "bare" description of the electronic
structure, when ignoring the weak exchange and correlation effects
implemented in present day DFT functionals.%
\footnote{In our calculations of the DOS, see Fig.\ 3,
we have used that of Perdew {\it et al.} \cite{Perdew1992,Perdew1996}.}
\begin{figure}[b]
  \includegraphics[width=0.42\textwidth]{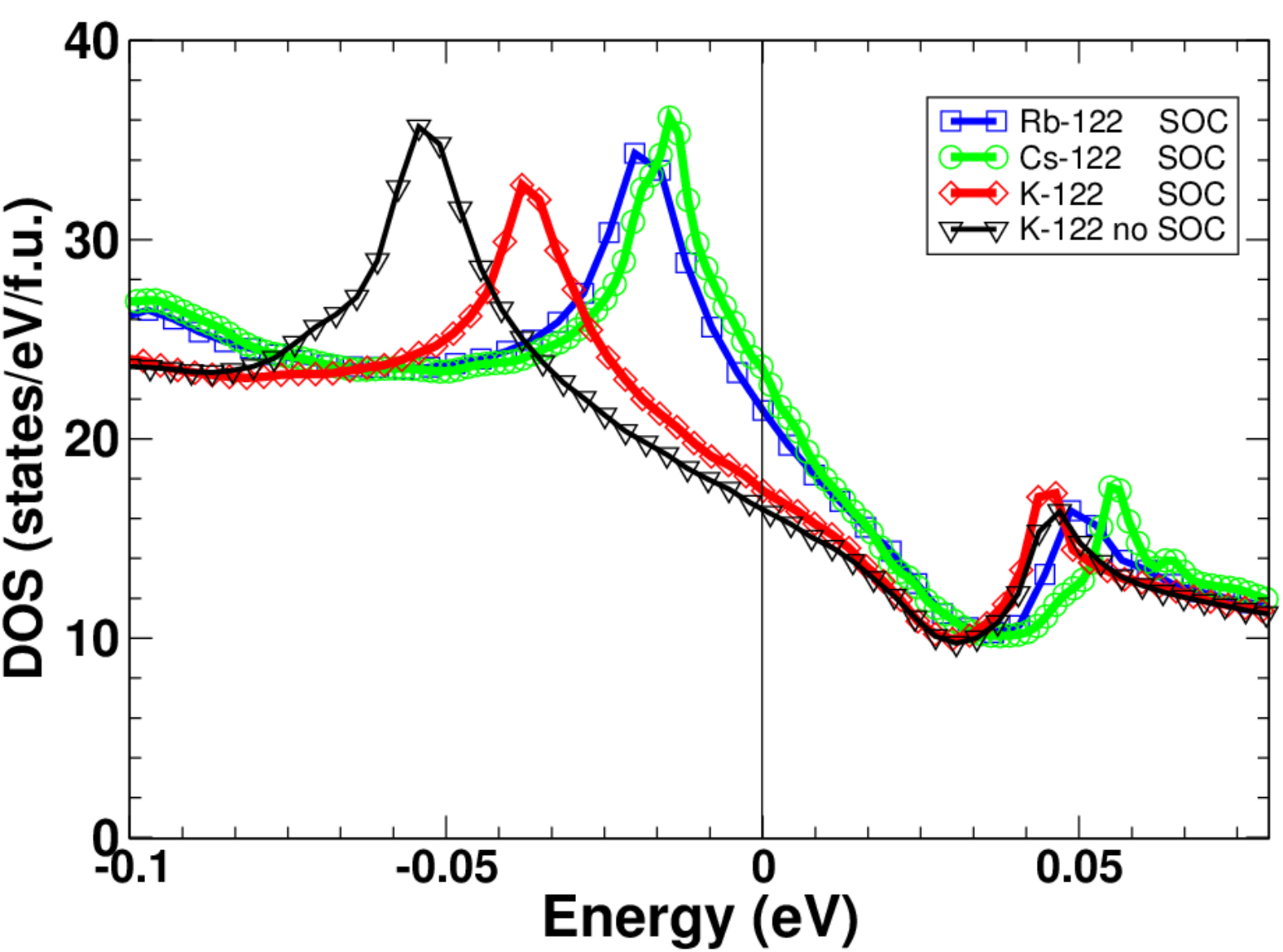}
\caption{(Color online) DOS of A122 A= K, Rb, Cs within the
quasi-particle self-consistent (QS)$GW$\cite{Kotani2007} approximation.
The spin-orbit coupling is added perturbatively after 
self-consistency is reached. See Ref.~\cite{Tomczak2012,Tomczak2015}
for results concerning other pnictides. Notice the puzzling relative close vicinity
of the 3$d_{xz}-3d_{yz}$ derived VHS position at -14~meV close to our ARPES data
at -11 meV at variance with the K122 data.}
\label{fig:3}       
\end{figure}
\begin{figure}[t]
  \includegraphics[width=0.42\textwidth]{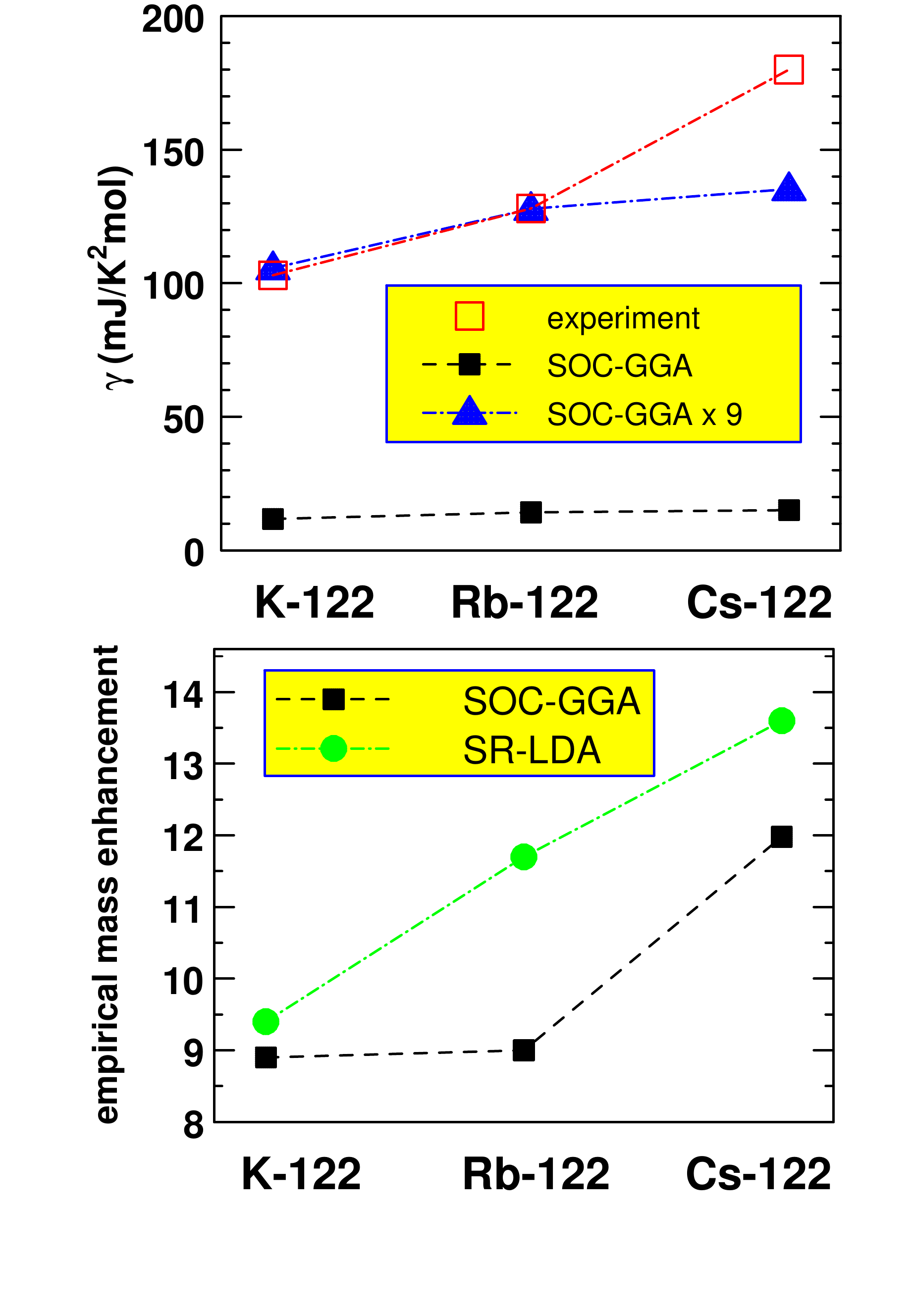}
\vspace{-1.0cm}  
\caption{Sommerfeld constants $\gamma$ vs.\
bare DOS from DFT.}
\label{fig:4}       
\end{figure}
There
are two observations providing direct evidence for significant
many-body effects 
in the strongly hole-doped end members
of the so-called 122 compounds $A$Fe$_2$As$_2$ with $A=$~K, Rb, Cs:
(i) the mass
enhancement of various bands crossing the Fermi energy and (ii) a
shifting of these bands resulting in different filling factors.
The latter effect is sensitive to approaching
a filling ratio closer to half-filling for the most strongly correlated
Fe-3$d_{xy}$ states which form the outermost Fermi surface sheet 
(FSS) centered around the $\Gamma$-point of the
Brillouin zone (BZ). The el-el interaction
affects also the position of various orbital-dependent 
VHS: In the present case they stay
closer to the Fermi energy by a factor of 3 as compared to the best DFT predictions,
(see Fig.\ 3).
Here we will focus briefly on the closest to $\varepsilon_{\rm \tiny F}$
Fe 3$d_{xz}$/$d_{yz}$ VHS which occurs in the middle of the $\Gamma$-X
line in the BZ. The latter is the center of the blade/propeller-like
$\varepsilon$-FSS \cite{Drechsler2017,Yoshida2014,Hardy2013,Hardy2016} 
clearly visible in Fig.\ 2.  
The corresponding band has been suggested to
dominantly bear superconductivity, although, both,
nodeless \cite{Hardy2013,Hardy2016} and  
$d_{x^2-y^2}$-nodal type \cite{Drechsler2017,Efremov2017,Hafiez2012,Grinenko2014,Reid2012}
characters have been proposed.

\section{Thermodynamic puzzles}
We start with an analysis of the available
specific heat data shown in Fig.\ 5. Using our bare DOS  $N(0)$ 
GGA-derived values
shown in Fig.\ 3, one arrives at a total mass enhancement 
of about  8.9 for K122, rather similar to about 9 for Rb122,
whereas in Cs122 an enhanced value $\approx$ 12 is realized \cite{Drechsler2017}. 
With the low-energy bosonic weak coupling constant $\lambda \approx 0.7$
taken into account, one is left with a dominant high-energy 
renormalization exceeding five. With the same or slightly reduced
high-energy renormalization, but enlarged low-energy coupling, the 
Cs122 data can be understood in qualitative accord with the somewhat
 reduced shift of the VHS considered in Sect.\ 5 dominated
 by the high-energy interactions $U$ and $J$ (see below).
The magnetic susceptibilities
$\chi(T)$
of K122 and Cs122 show a broad maximum at low $T_{\rm \tiny max}$.
This has been considered as the hallmark for strong correlations ascribing  it
to the freezing temperature $T_f$ generic for bad metals with well-defined 
quasi-particles and Fermi liquid-like behavior below $T_f$, only.
Then the observed systematic downshift of $T_{max}$ on going from
K122 via Rb122 to Cs122 could be a signature of increased correlation
strength \cite{Hardy2016,Wu2016}. However, our experimental data  
for the less strongly 
hole-overdoped samples, 
exhibit only
a very moderate downshift and an anomalous 
behavior of the Wilson ratios $R_{\rm \tiny W}$ (see Figs.\ 6 and 7). This 
points to a more complex scenario,  probably related to the multiband nature
of these compounds with rather different correlation regimes in the various subsystems.
Indeed, our low-energy bosonic scenario \cite{Drechsler2017} 
mentioned above could also be helpful in resolving these
puzzles related to the smaller $\chi(0) $ of Cs122 in comparison
to Rb122 and K122, as well as to their Wilson ratios $R_W$.
Noteworthy, a large and even an increased 
\begin{equation}
R_W= \frac{4\pi^2k_{\mbox{\tiny B}}\chi(0)}{3g^2\gamma (0)} ,
\end{equation}

\begin{figure}[t]
  \includegraphics[width=0.4\textwidth]{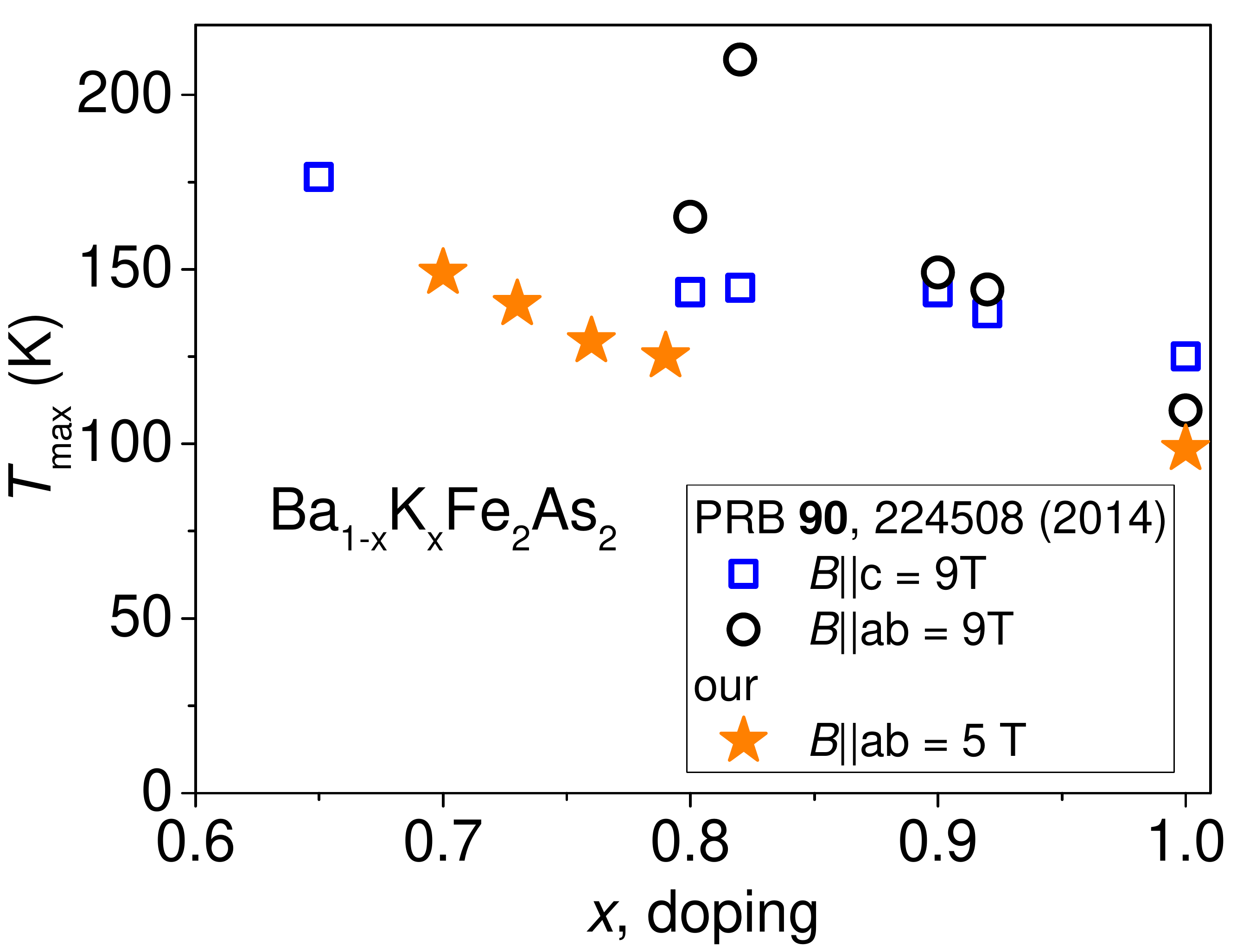}
\caption{Low-$T$ maximum of the magnetic susceptibility 
in hole-overdoped K-doped Ba-122 single crystals  vs.\ doping ratio.
from our data and from those of Liu {\it et al.} \cite{Liu2014}.}
\label{fig:2}       
\end{figure}
\begin{figure}[b]
  \includegraphics[width=0.4\textwidth]{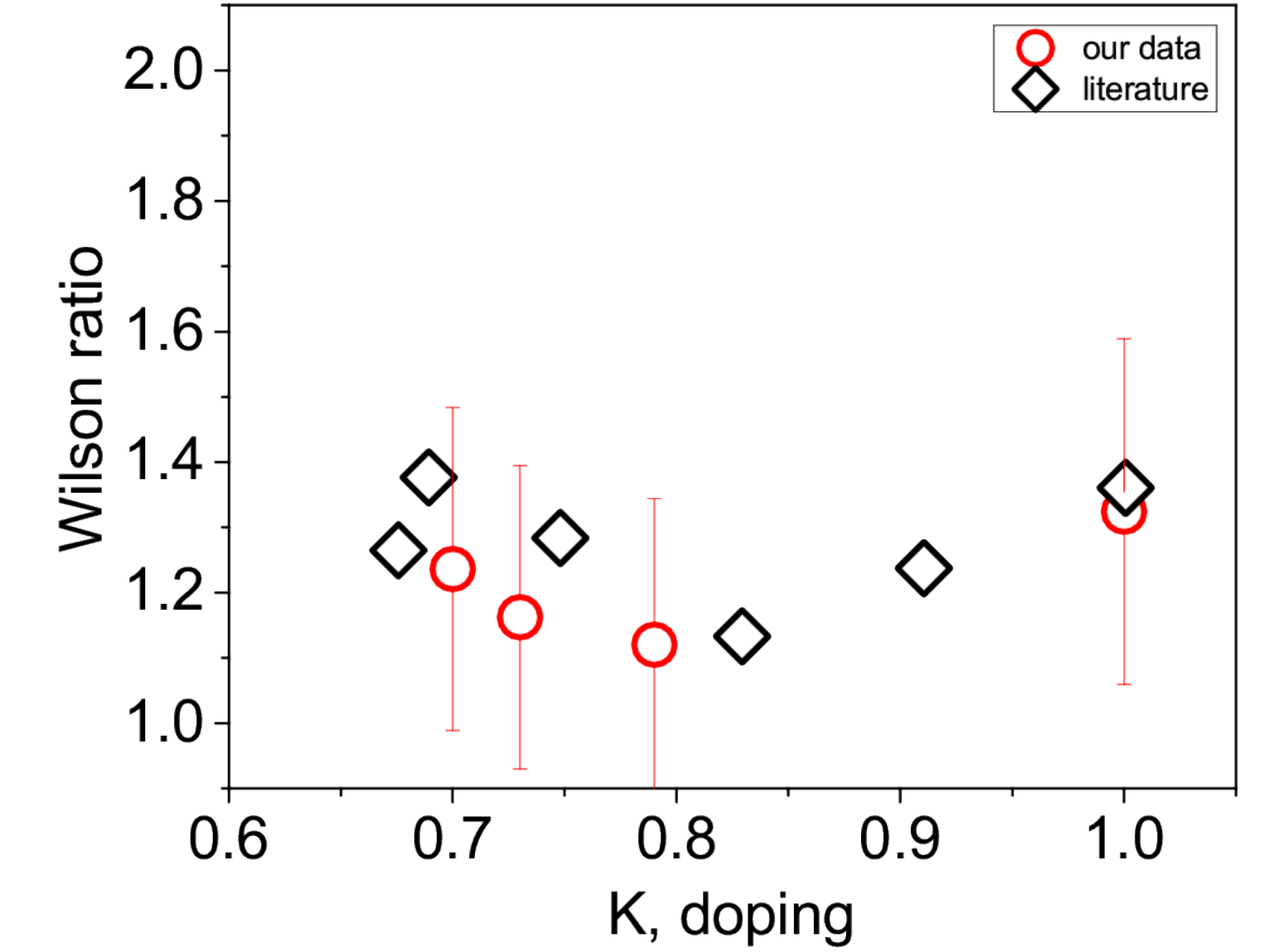}
\caption{Wilson ratio in h-overdoped K-doped Ba-122 vs.\
doping from our data and that of Refs.~\cite{Liu2014,Hardy2016}.
	Notice the moderate 
  but non-monotonous dependence with a minimum near the
  the Lishitz point where the two electron pockets near the corner of the BZ
  disappear, and the seemingly
  weakly correlated values well below 2.} 
\label{fig:6}      
\end{figure}
is widely used as a direct measure of the relative strength of correlation
effects.  Starting from $R_W$({\small K-122}) $\approx 1.35$ 
taken from Fig.\ 3, and using the data from Ref.\
\cite{Wu2016} (Fig.\ S4 and Tab.\ S1 therein), one estimates 
$R_W$({\small Cs-122})/$R_W$({\small K-122})$\leq $~0.394,  while  less
correlated FeSC exhibit $2 \leq R_W \leq 5$.
This is the first
counter-intuitive observation. Since
this values is in
obvious conflict with the notion that stronger correlations are present in
Cs122 and Rb-122 when compared to K122.\footnote{ We mention that among a
large number of known (until 2000) heavy-fermion superconductors,
only three of them (CeCu$_2$Si$_2$, UPt$_3$, and UBe$_{13}$) exhibiting 
surprisingly formally ``free electron"
Wilson
ratios $R_W \approx 1$ \cite{Radousky2000}.}  To the best of our knowledge,
nowhere such a low $R_W \sim 0.4$ has been observed as in Cs122, 
yielding the 2nd counter-intuitive observation.
The position of the VHS 
reported below points to a weaker correlation regime in Cs122
than in K122, like in Ref.\ \cite{Backes2015}.
\vspace{-0.3cm}
\section{Van Hove singularities as seen by ARPES}
\label{sec: 5}
To determine the position and the correct orbital nature of the VHS
of both
single crystals mentioned in Sect.\ 2,
angle-resolved photoemission spectroscopy (ARPES) data have been collected 
at the I05 beamline of Diamond Light Source. Single-crystal samples were 
cleaved {\it in situ} in a vacuum better than 2$\cdot 10^{-10}$ mbars and measured 
from 5.7 to 270~K. These measurements were performed 
using linearly polarized synchrotron light, utilizing
a Scienta R4000 hemispherical electron energy analyzer with an angular 
resolution of 0.2 to 0.5$^o$ and an averaged energy resolution of 
about 3~meV. 
(see Fig.\ 2 for the case of K122).
The position of the Fe 3$d_{xz}/3d_{yz}$ VHS as derived from our ARPES 
measurements
of about -$15 \pm 1$~meV and -11$\pm 1$~meV for K-122 and Cs-122, respectively. 
\footnote{Note that our results 
differ quantitatively by a systematic shift of nearly 10 meV for the entire
spectrum as compared to Ref.\ \cite{Fang2015}, and even qualitatively in that previous 
Cs122-STM 
\cite{Yang2016} showed no VHS at all.
 We speculatively ascribe this  to surface reconstructions and charging
 in those samples most detrimental for STM. The position in STM of the VHS
 was given as -5 to -6 meV for K122. },
 respectively, significantly
differs as expected from the DFT values by a factor of four which clearly points
to strong many-body effects. Also the {\it GW}-approximation in the present
form \cite{Kotani2007,Tomczak2012} does not yield a good description, caused
probably by overscreened Coulomb interactions. Its future combination
with a DMFT-like approach \cite{Tomczak2015,Tomczak2017} might resolve this
problem. We admit that a fully relativistic GGA-based code might 
reduce the deviations from the ARPES data.
The order of magnitude of the VHS-shifts can be approximately
reproduced by the slave spin and/or the DMFT theory using appropriate
phenomenological parameters for the Hubbard $U= 2.7$ to 4~eV
and the Hund's exchange $J=U/4$ \cite{Medici,Backes2015}.
Both are responsible also
for the high-energy renormalizations for the 
DOS and the Sommerfeld coefficient $\gamma$. In general, one may expect 
the following decomposition of the VHS-shift relative to standard non-relativistic DFT 
\begin{equation}
\delta \mu_{\rm \tiny VHS} =\delta \mu_{\rm \tiny so}+
  \delta \mu _{\rm \tiny high}(U,J)+
  \delta \mu _{\rm \tiny low b} \quad .
 \end{equation} 
 where the three terms stand for the SO coupling, the
 high- and low-energy bosonic contributions, respectively. According to 
 Fig.\ 2, and Ref.\ \cite{Drechsler2017}
  the 1st terms amount to $\approx$18.3 (12.2, 8.75)~meV for 
  K(Cs,Rb)122, respectively.
Then one is left with empirical
 shifts  of many-body origin of
$\approx$ 45~meV for K122 and 37~meV for Cs122, yet to be understood theoretically.
The 2nd term in Eq.\ (2) is the high-energy contribution.
It should scale with the matrix element of the 
high-energy 
self-energy $\Sigma_{\rm el-el,h}$
$\propto (U+ \alpha J)^2$ where $\alpha$ is a value in between
-1 for Fe$^{2+}(d^6)$
and $\alpha =+ 4$ for full 3$d$ Mott systems at half filling 
(i.e.\ at Fe$^{3+}(3d^5)$) realized in some Mn or Cr systems. 
For simplicity we adopt $\alpha =1.5$ 
and an often used $J=0.25U$ ratio. 
Anyhow, it should be nearly the same for K122 and Cs122 
and will therefore not affect our main conclusion.
This contribution reads 
\begin{equation}
\mu _{\rm \tiny high}(U,J)\propto (U+ \alpha J)^2\approx1.89U^2 ,
\end{equation}
and provides a convenient measure of the effective Hubbard interaction
$U_{eff}=U+\alpha J$ or of its
relative change from K122 to Cs122 at the same doping but
at fixed ratio $J/U=0.25$ . It 
is expected to be dominant
since the low-energy bosonic contribution should be much weaker, although
not necessarily the same
 for a coupling to soft critical modes and this way different for the three
compounds (see below). 
Ignoring it to first approximation and using the empirical absolute shifts,  
one would arrive at 
a ratio of about 1.216 or at 
$U_{\rm \tiny K122} \approx  1.103U_{\rm \tiny K122}$, i.e.\ at {\it weaker} correlations
for Cs122 in accord with the reduced $\chi$ and the
Wilson ratio, but in sharp contrast to opposite claim of
Ref.\ \cite{Eilers2016} where an {\it increase}
 of $U$ of $\approx$ 20\% has been suggested for Cs122.
 A reduced $U$-value for Cs122 might be due to the larger
 polarizibilities of the larger Cs$^{1+}$ cations (by $\approx$ a factor of 3)
 as compared to 
 K$^{1+}$, leading to more screened $U$-values.
 Note that a significantly higher energy of the VHS by $\approx$ 8 meV
 i.e.\ at -3~meV 
 \footnote {Beyond the error bars of about 2 meV for sharp peaks.}  for Cs122
 would be required to arrive at the same effective $U$
 in the present estimate. In the scenario of Ref.\ 3
 the VHS
 would even occur  {\it above} or {\it at} $E_{\rm \tiny F}$. 
 
 Indeed, adopting an effective two-band model
 by dividing the total system into one subsystem containing
 the Fe-3$d_{xz}/3d_{yz}$ electrons (which produce the VHS considered here) and a second for the remaining
 part of electrons, one might estimate 
 the bosonic shift for the chemical potential
 following Ref.\ \cite{Abrikosov1962} as
 \begin{equation}
 \delta \mu _{\rm \tiny low, b} \propto \lambda_{\rm \tiny b}\omega_{\rm \tiny b} .
 \end{equation}
 Using $\lambda_{\rm \tiny b, K122} = 0.7 $ \cite{Drechsler2017}
 and $\hbar\omega_{\rm \tiny b}=7$~meV from inelastic neutron scattering 
 (INS) data
 \cite{Lee2012}.
  The order of the experimental difference  of $\approx 4 \pm$ 2~meV
 can be readily understood, if a stronger coupling 
 to low-energy bosons
 for Cs122 due to the 
 closer vicinity to a 
 QCP \cite{Eilers2016}  (see Fig.\ 1) is taken into account.  
 The incommensurate spin fluctuations seen in the INS \cite{Lee2012} 
 and NMR data for Cs122 \cite{Zhang2017} point
 to a phase different from the AFM commensurate Mott one.
 Measurements of the VHS for Rb122 and other Fe-pnictides
 are highly desirable.
\section{Summary and outlook}
Shifts of van Hove singularities relative to DFT-predic-
tions provide a direct measure of the high-energy
el-el interaction. There is no obvious increase of the 
Hubbard $U$ and the Hunds's exchange $J$
ongoing from K122 to Cs122 by analyzing phenomenologically
the 3$d_{xz}-3d_{yz}$ VHS (closest to $\varepsilon_{\rm \tiny F}$).
Its shift is affected by the spin-orbit coupling which cannot be ignored
(as done previously in the literature) for a reliable estimate of many-body effects.
For a full description, also the VHS  at larger binding energies
related to the Fe-3$d_{xy}$ and the 3$d_{z^2}$ orbitals 
should be studied experimentally. 
More results will be given elsewhere. In general, a  
combined theoretical and ARPES
study
of VHS shifts provides a valuable general tool to study many-body effects
in quasi-2D  systems in addition to those reflected in the mass enhancements.
\acknowledgement
S-LD, SW, SA, LM, and AB 
thank the Volkswagenstiftung for financial support.
IM acknowledges support  by the 
RSF grant No.\ 16-42-01100. VG and SB
are 
grateful to the DFG for financial support through the grants GR4667 and
 BO1912/6-1 and BO1912/7-1, respectively.
 We acknowledge the Diamond Light Source for time on the 
 i05-ARPES beamline under the Proposal SI96989.
Discussions
with M.\ Daghofer, D.\ Efremov, J.\ Fink, E.\ Goremychkin,  and L.\ d\'e Medici are 
gratefully acknowledged.\\

\noindent
{\it \bf Note added for the readers of arXiv}\\
Within a very recent preprint: arXiv:1807.00193v1 (submitted 30 June 2018)
devoted to a chemical pressure tuning of the van Hove singularities in
KFe$_2$As$_2$ and CsFe$_2$As$_2$ revealed by an ARPES study
by
P.\ Richard ,
A.\ van Roekeghem, X.\ Shi, P.\ Seth, T.K.\ Kim, X.-H.\ Chen, S.\ Biermann, 
and H.\ Ding, 
very similar positions of the saddle-point van Hove singularities
have been found: -15~meV to -12~meV as well as -10~meV as compared to 
-15$\pm$ 1~meV  
and -11$\pm$ 1~meV,
respectively, reported above.

\end{document}